1st Yang Yue, *Institute of Science Studies and S&T Management, Dalian University of Technology*

*Doctor candidate, No. 2 Linggong Road, Ganjingzi District, Dalian City, Liaoning Province, P.R.C., 116024*

*Institute of Science Studies and S&T Management, Dalian University of Technology*

[ysynemo@mail.edu.dlut.ed.cn](mailto:ysynemo@mail.edu.dlut.ed.cn) *18104085833,微信：@y1lbw2013*

2nd Joseph Z. Shyu, *Institute of Science Studies and S&T Management, Dalian University of Technology*

*Professor, No. 2 Linggong Road, Ganjingzi District, Dalian City, Liaoning Province, P.R.C., 116024*

[joseph@cc.nctu.edu.tw](mailto:joseph@cc.nctu.edu.tw) *微信@jzskf1123*

# An overview of research on human-centered design in the development of artificial general intelligence

Yang Yue [1st]   Joseph Z. Shyu [2nd]

*Institute of Science Studies and S&T Management, Dalian University of Technology*

ysynemo@mail.edu.dlut.ed.cn ， joseph@cc.nctu.edu.tw

**Abstract:** This paper provides an overview of the development of Artificial General Intelligence (AGI) from a humanistic perspective. Drawing upon a range of sources, including academic literature and industry reports, the paper examines the evolution of AGI and the key technological and ethical challenges that are associated with its development. In particular, the paper focuses on the ways in which AGI may impact society and individuals, and the importance of ensuring that AGI is developed in a way that is aligned with human values and interests.

**Purpose**

This paper provides an overview of research exploring the role of human-centered design in the development and governance of general artificial intelligence.

**Design/methodology/approach**

The research methodology used in this article includes content analysis and literature review. Content analysis is used to examine and analyze the main themes and concepts related to human-centered design in the development of general artificial intelligence. Literature review is used to identify and analyze relevant research studies, theories, and best practices related to this topic.

**Findings**

Human-centered design is crucial for the development of general artificial intelligence (AGI) that respects human dignity, privacy, and autonomy. The integration of values such as empathy, ethics, and social responsibility in the development and deployment of AGI can lead to more ethical and sustainable use of the technology. Talent development for AGI is crucial to meet the needs of the field, and interdisciplinary approaches can be used to cultivate "general, intelligent, and versatile" talents.

**Research limitations/implications**

Lack of further empirical research is needed on human-centered design in the development of AGI, particularly in the areas of ethics, social responsibility, and talent development.

**Practical implications**

By incorporating human-centered values such as empathy, ethics, and social responsibility into the development and deployment of AGI, it is possible to achieve more ethical and sustainable use of the technology. This can lead to the responsible and ethical use of AGI, which respects human dignity, privacy, and autonomy. Furthermore, it is essential to strengthen talent cultivation for strong

AI, which is the foundation of scientific research. Through collaborative innovation among industry, academia, research, and application, a solid talent guarantee can be provided for the healthy and stable development of AGI.

**Originality/value**
The originality of research lies in its exploration of the importance of a human-centered approach to the development of AGI. The paper highlights the need to integrate human-centered values into the design and deployment of AGI. It also discusses the practical implications of this approach.

**1. introduction**

The 21st century is an era of explosive technological innovation, with the emergence of artificial intelligence, Internet of Things, blockchain, and even quantum technology. Humanity is approaching the singularity at an infinite pace. AGI has taken its initial shape, and the emergence of AIGC, represented by Chat-GPT3.5, has led to a great explosion and surge of creativity for humanity. As of now, artificial intelligence (AI) has become a transformative technology that is rapidly changing every aspect of human life. Humanity is rapidly investing in the development of Artificial General Intelligence (AGI), also known as Strong AI, which is an advanced form of AI designed to create machines that can perform any intellectual task that a human can do. AGI has the potential to revolutionize many fields, including healthcare, education, finance, and transportation. If 2018 was the year when Artificial General Intelligence (AGI) became mainstream, then today, five years later in 2023, will be a year of explosive development in artificial intelligence.

The emergence of ChatGPT at the end of 2022 has profound significance, as the global technological revolution approaches singularity. As a large-scale language model in the field of Artificial General Intelligence Collective (AIGC), ChatGPT has become a bridge for AI to move towards AGI. Human language is the basis for communication between people, and ChatGPT has enabled machines to understand and generate natural language, achieving smooth communication between humans and machines. It enables machines to handle various language scenarios accurately and efficiently, and gives machines human-like response capabilities. This can be seen as a precursor to humans achieving general artificial intelligence. Although current ChatGPT still cannot match humans in terms of computational power and emotional intelligence, its capabilities already exceed the average level of humans and are advancing towards higher goals with continuous development, enabling machines to have human-like thinking abilities.

From what we can see at present, the potential of AGI exceeds human learning abilities, and rethinking the relationship between humans and AGI has become a hot topic this year. Humanism is once again entering people's field of vision. In the process of human production and life, we can clearly feel that tools and technology exist as tools for human activities. Whether technology is the first productive force or activity theory, AGI artificial intelligence is a typical representative of technological revolution. Not only the natural language model of AIGC, but also the initially designed general artificial intelligence (AGI) with human-like thinking abilities in the future, are all aimed at improving the quality of human production and life. In essence, they are all centered around humans. Just like how the Industrial Revolution liberated human physical labor through machines, we can understand AGI as the liberation of human productive power at the level of physical and mental labor through tools. When AGI succeeds, it will be a milestone for humanity to transcend its biology.

In recent years, the development of artificial intelligence (AI) has been a crucial area of research and innovation. As technology continues to advance, it becomes increasingly important to consider the role of human values and perspectives in guiding its development. As AI permeates various industries and aspects of daily life, it is crucial to consider its development from a human-centric perspective. Human-centered values such as empathy, creativity, and moral responsibility can help shape the direction of AI development, ensuring that it benefits society and remains consistent with human values. However, there is a lack of consensus on what these values and perspectives should be and how to integrate them into AI development.

This paper aims to explore the role of the human-centric perspective in promoting the development of artificial general intelligence (AGI) by answering the following research questions: (1) What is the current status of AGI development? (Chapter 2) (2) What are the human-centered values and perspectives, how are they understood domestically and internationally, and what is their relationship with AGI development? (Chapter 3) (3) What are the challenges and opportunities of promoting AGI development from a human-centric perspective? (4) How can policy makers ensure responsible development and use of AGI from a human-centric perspective?

The development of AGI must follow an ethical framework that prioritizes human values, in order to bring about positive impacts on human life, the economy, and society. The human-centric perspective provides a reference framework, emphasizing the importance of human values as active agents in the process of technology governance. This study aims to provide a comprehensive review of relevant literature on AGI development from a human-centered perspective and explore the role of human-centric values in promoting AI development. The Web of Science academic database was selected, and the search terms used were "people-oriented" OR "human-centrality" AND "AGI" OR "AI". The content analysis method was used to interpret the literature and identify key themes and trends in the field.

The purpose of conducting a review of the development of artificial general intelligence (AGI) from a humanistic perspective is to explore the role of human values and viewpoints in guiding its development. As AI continues to penetrate various industries and aspects of daily life, it becomes increasingly important to consider the impact of its development from a human-centered perspective. Human values such as empathy, creativity, and moral responsibility can help shape the direction of AGI development to ensure it benefits society and aligns with human values. However, there is currently a lack of consensus on what these values and viewpoints should be and how they can be integrated into the development of AGI. The review aims to address this gap by answering key research questions related to the role of humanism in AGI development, and to provide guidance for policymakers and industry professionals on how to responsibly develop and use AGI in line with human values.

## 2. Machine-Centered and Human-Centered of AGI Development

Artificial General Intelligence (AGI) technology aims to enable machines to think like humans and perform a variety of tasks through capabilities such as transfer learning. AGI is currently at the forefront of AI research and development. Unlike the human brain, AGI's computing and analytical abilities are theoretically limitless, with efficient data collection, processing, and analysis capabilities, as well as cognitive and learning abilities similar to humans. AGI is a general-purpose artificial intelligence system capable of performing various tasks. AI originated from academic research in the 1960s and has shown tremendous potential only in recent years with the support of large amounts of data and computing power, exhibiting rapid explosive growth.

AGI is the ultimate goal of the field of artificial intelligence, referring to machines that can perform human-like intelligent tasks such as reasoning, problem-solving, and learning. In recent years, research on AGI has shown a rapid increase in the number of publications and research results, as shown in Figure 1. Scientific quantitative analysis of AI and AGI-related research in recent years using literature collected from WOS indicates that the total number of AI-related publications has dramatically increased since 2012. In addition to data, new hardware innovations in computing power such as quantum computing and optical computing may also bring about changes and breakthroughs in AGI. At the same time, there are gaps in computer fundamental theories that need to be filled. Currently, existing representation methods are not suitable for identifying tasks, and AGI requires a deep understanding of time, space, and causality.

AGI is a research focus in the field of artificial intelligence and is generally regarded as the ultimate goal of AI development. Just yesterday, on March 16th, OpenAI released GPT-4, a large multimodal model developed on the basis of GPT-3.5, which can handle image and text input and generate text output, such as dialogue systems, text summarization, and machine translation, with extremely powerful natural language processing capabilities. It has even achieved very high scores in various human-designed professional exams. Compared to the concept of AGI, it has already reached a very high level and is close to domain-specific AGI. GPT's performance in academic and professional exams is impressive, with simulated scores approaching or even exceeding the average human exam scores. As former OpenAI founder Elon Musk emphasized, with more advanced computing power, AI development will accelerate, and AI will become an issue that humans cannot avoid in the future.

Despite this, there is still some way to go before humans reach the level of true general artificial intelligence (AGI), and currently AGI is in the stage of artificial narrow intelligence (ANI). The concept of AGI was proposed by American philosopher Searle in 1980, pointing out the development path of artificial intelligence. Early artificial intelligence researchers include Marvin Minsky, Allen, Newell, Herbert Simon, John McCarthy, etc. The process from weak AI to strong AI is a necessary stage for achieving AGI. Strong artificial intelligence (AGI) can process various intelligent behaviors through a set of systems, while weak artificial intelligence requires new independent systems for each intelligent behavior, which is the fundamental difference between the two. Early AI research mostly followed the "knowledge-based" approach, sometimes referred to as GOFAI (Good Old Fashioned AI). A complete AGI system needs to be able to solve global problems (global properties), such as the ability to select which theoretical system is more suitable for the current application scenario, and the ability to be concise and consistent with existing knowledge systems. This ability is defined as the ability of AGI systems to mimic human reasoning and problem-solving abilities, and even to have independent self-awareness. Current AI technology obviously cannot meet this standard, and the products provided by mainstream AI research do not belong to the category of general intelligence. Professors Marcus (Gary Marcus) and Davis (Ernest Davis) from New York University proposed in 2019 to rebuild artificial intelligence systems that we can trust. With the advent of quantum computing, this exploration has become possible. Due to the epoch-making breakthrough in computing power, another approach compared to deep learning is to study brain-like mechanisms, and the current popular research approach of "brain-inspired AI" is a manifestation of this idea. By mimicking the working mode and principles of the human brain, it is possible to achieve general artificial intelligence in the near future.

The development of the artificial intelligence industry supported by indigenous technology is of great significance for China. Focusing on core scientific issues, encouraging fundamental and cross-disciplinary exploration, and prioritizing original research are all key aspects of the country's approach. China is prioritizing the development of innovative methods for the next generation of AI, which will promote related research and applications, develop new algorithm models, databases, knowledge bases, etc., and promote the development of AI in various fields. In the past two years, China's emphasis on AGI has been no less than that of the West. Through the collaboration of the government and universities, China is developing a new method system for AI and promoting fundamental research and talent development, which will support the country's dominant position in the new round of international scientific and technological competition. In the industry, AGI is experiencing explosive development, with a reported 15.2 billion yuan invested in AI start-ups in 2017.

There are multiple ways to develop artificial intelligence (AI) from a human-centered perspective. One approach is to focus on developing AI that can mimic human intelligence and behavior, emphasizing the importance of understanding human cognition and behavior to create intelligent systems that interact with humans in a more natural way. Another approach is to develop AI that can enhance human capabilities and well-being, supporting activities such as healthcare, education, and entertainment. A third approach is to develop AI that reflects human values and ethics, emphasizing the importance of embedding human values and ethics in the design and development of AI systems to ensure they align with human interests and needs.

The tension between machine-centered and human-centered approaches to AI development. Developing AGI is a complex and challenging task that requires the integration of different disciplines such as computer science, mathematics, neuroscience, cognitive psychology, and philosophy. In recent years, developing AGI from a human-centered perspective has become increasingly important, as people recognize the need to align AI with human values, ethics, and social responsibility. Human-centered principles guide the development of AGI by reflecting human values, preferences, and needs at the design level to ensure that AGI's goals and behavior align with human values. The concept of value alignment has been proposed to incorporate ethical and social considerations into the development and implementation of new technologies, emphasizing the need to align technology with human values and ethics.

The ideas of machine-centered and human-centered approaches to AI development stem from the most basic starting points of human beings' pursuit of technological progress and innovation. Vannevar Bush believed that science was not omnipotent and advocated that the direction of basic technological research must be strategically guided by the government, which is a typical optimism or technology innovation mentor. On the contrary, technology enthusiasts such as Harley Kilgore believed that technology can solve anything, which is a typical pragmatism. In fact, in the field of technological innovation, there are two main schools of thought (or methods). One is called "instrumentalists," which is the thinking of the supply side and is also for the pursuit of universal truth; the other is called "realists," which is the thinking of the demand side and is also for the pursuit of substantive technological competitiveness, and belongs to the metrology school for the former and the practical school, which is the technological system that Einstein called "visible" or "measurable," for the latter. In the STEM technology architecture, obtaining a common scientific understanding through either metrology or practice is the core of modern digital economy and technological innovation. The two factions of AI development that have emerged from this are the machine-centered artificial neural network school and the human-centered neuromorphic computing school, each with its own characteristics, advantages, and disadvantages.

**2.1 Machine-Centric Artificial Neural Network**

Machine-Centric Artificial Neural Network focuses on simulating the structure of the human brain. The most well-known example of this approach is deep learning, which is a type of artificial neural network that simulates the human brain. Witnessing a computer program mastering the most complex human games to a world-class level in just a few hours is an awe-inspiring experience, as it approaches a form of intelligence. Computer science-based approaches emphasize simulating human external behavior through machine programming, without delving into the underlying driving forces, which is typical of symbolic and behavioral research methods. AGI aims to replicate not only human learning ability, but also all human behavior.

The machine learning field mostly focuses on "supervised" learning, which requires a large number of labeled samples to train machines to recognize similar patterns. Deep learning operations typically involve a large number of training samples and consume significant computing resources. For example, in image recognition, a recognition task is set, a hierarchical neural network architecture is designed, and specific information aggregation and activation functions, the number and arrangement of calculation units, and the property types that each layer of calculation units must extract, as well as the feedback algorithm for the entire framework, must all be pre-set. Moreover, a large number of image samples are required for training.

Thus, developing a deep learning system that can cross multiple domains will largely involve the thorny issues of "where do training samples come from" and "where do computing resources come from?" In contrast, humans do not face such problems and can acquire new skills in a new field with very few training samples. Therefore, some scholars believe that simulating human brain logical functions could provide insight into the development of AGI.In the field of unsupervised learning, transfer learning, which involves migrating a training model to another task, has gained increasing attention in recent years. This approach can significantly reduce model training time and improve model reusability in applications. In addition, unsupervised learning is also critical, but currently cannot meet the standards of AGI.

Although deep learning-based AI methods have made important breakthroughs in many scenarios, the current machine learning and AI are still plagued by problems such as poor model interpretability, weak robustness against adversarial samples, high demands for data and computing power, and weak theoretical foundations. Supervised and unsupervised learning together constitute the basic logic of weak AI based on deep learning, which is currently the hottest field in machine learning and AI.

The viewpoint supported by this faction is that general artificial intelligence (AGI) can be achieved gradually through the accumulation of specialized artificial intelligence technologies such as image recognition. However, they ignore the difficulties caused by industry segmentation. The current mainstream deep learning technology needs to face a problem if it wants to achieve the standard of AGI in the form of stacking blocks, which is that deep learning frameworks specialized in facial recognition cannot be directly used for playing Go, and deep learning frameworks designed for playing Go cannot be directly used for playing Chinese chess. Therefore, a deep learning system that can span more fields of work will involve the thorny issues of "where the training samples come from" and "where the computing resources come from". Compared with this, humans can often learn new skills in new fields by analogy even when training samples are scarce. American biostatistician Jeff Leek pointed out that deep learning technology will become a "dragon-slaying technique" unless you have a massive amount of training data. The basic principle of deep learning is to explore the model's dependence on hyperparameters, understand the principles behind the model, and establish approximation theory, generalization error analysis theory, and convergence theory of optimization algorithms for deep learning methods.

Since all true AGI systems should be able to switch working knowledge domains according to changes in tasks and environments under unsupervised conditions, it is difficult for deep learning systems to evolve into AGI systems no matter how they develop. AGI systems must have human-level common sense reasoning ability, and a basic characteristic of common sense reasoning is that the domain involved in the reasoning process cannot be determined in advance. If we hope that AGI has human-level problem-solving ability in diverse problem domains, the agent must have the ability to globally balance requirements from different domains, which is the ability to solve problems with global properties. The above requirements are not only imposed on humans but also on an ideal AGI system. AI researchers believe that unsupervised learning, including autoencoders, deep belief networks, and GANs, is key to achieving AGI. The challenges of achieving general artificial intelligence through deep learning systems and the importance of unsupervised learning in developing AGI.

**2.2 Human-Centered Neuromorphic Computing**

The human-centered neuromorphic computing approach, which is based on simulating human brain logic, is a school of expert systems that is guided by neuroscience and driven by human behavior. It employs neural networks to achieve neuropsychological and cognitive behavioral functions, with the core focus being the study of the biological brain's neural operating mechanisms. Its aim is to enable machines to simulate the human brain as accurately as possible, embodying the paradigms of connectionism and functionalism. This approach believes that by accurately describing the details of the human brain's neural operation, a precise blueprint for AGI can be abstracted from it. While artificial neural networks in machine learning are also based on biomimetic simulation of the human brain, in the eyes of professional neuroscientists, both traditional neural networks and more complex deep learning mechanisms are still very primitive and localized in their simulation of the brain. Considering the overall operating mode of the human brain, rather than just the operation of localized neural networks, provides us with more information about the overall operation of human intelligence. Therefore, research on neuromorphic AI can obviously reduce the probability of "blind men touching the elephant" errors compared to mainstream deep learning research.

The "man is a machine" theory is the theoretical basis for AGI in this field. This theory believes that humans and machines have some essential similarities, so when humans understand how their own thinking works, they can create AGI based on the principles of the human brain. Currently, machines only exist in the von Neumann paradigm and do not have the cognition of the interrelation between different senses, only inheriting human cognitive results. For example, the biological neural network and the NCBL cognitive fusion system belong to this category. In this approach, the biological brain's neural operating mechanisms and cognitive behavioral mechanisms become the mainstream of research. Peter Gardenfors, a Swedish cognitive scientist, and others proposed the "cognitive spatial theory" of the brain, which holds that the hippocampus's place cells and grid cells not only map physical space but also concept space. In fact, some scholars abroad have attempted to develop a whole-brain connection architecture (WBCA) composed of information from experiential neural circuits throughout the brain, which has played various cognitive functions and behaviors similar to those of the brain in computational systems along this path.

Intuitive, non-logical and non-rational knowledge in human cognition can be transformed into logical and rational knowledge through processing, transformation, development, and sublimation. The goal of general artificial intelligence (AGI) technology is to achieve transfer learning abilities that insight into the essence of knowledge, which requires simulating human intuitive knowledge through a biomimetic pathway to achieve rational knowledge. The feasibility of AGI technology depends on whether a mechanism can be established to simulate human intuitive knowledge. Its objective is to create a general intelligent agent that has autonomous perception, cognition, decision-making, learning, execution, and social collaboration capabilities, while conforming to human emotions, ethics, and moral values. Based on the understanding that humans are machines, some of the logical thinking of humans is simplified into machine programs, enabling machines to achieve excellent computing levels in certain fields.

The biomimetic pathway of AGI technology is based on the cognitive process theory. From the perspective of the overall cognitive process, it emphasizes the determining role of practice in cognition. The brain consists of hundreds of components, each with its unique neural architecture, and they all perform their respective tasks in the operation of the entire brain. Neural networks are only partially driven by biology, but they are not computational neuroscience models. The OpenCog system can be considered biological to some extent, with nodes and neural connections, and it draws inspiration from neural networks. Although it does not replicate the brain at the level of neuroscience, the basic principles are similar. To have a universal framework for expression, many different algorithms are needed. Both the human brain and computers must have spatial analysis abilities, which are fundamental pattern recognition capabilities and cognitive patterns they face. This is a fundamental mathematical paradigm called the "geometry paradigm," which includes various branches of mathematics such as geometry, topology, and graph theory.

For example, the neuroscience pathway of general artificial intelligence and machine game refers to the use of the simulation of human brain thinking to construct a general intelligent machine. Based on the understanding that humans are mechanical, it simplifies some of the logical thinking in humans into machine programs, allowing machines to achieve superb computational levels in certain fields. If compared to the human brain, artificial intelligence can be seen as an extension of human intelligence outside of the constraints of living organisms. The breakthrough in the relationship between "biological" and "non-biological" intelligence, and the exploration of the general laws of intelligence, aims to make software systems have various cognitive functions such as perception, memory, emotion, reasoning, and decision-making, and even have learning abilities similar to human learners. General artificial intelligence will become a highly automated social mechanical general intelligence generated by a highly developed social brain. AGI is liberating the productivity of intellectual levels and is a milestone in which humans philosophically surpass biology (When Humans Transcend Biology).

However, there are also many doubts about this school of thought. Firstly, some scholars believe that deep structured learning inspired by information processing and distributed communication nodes in biological systems only symbolically simulates the static structure of the human brain. So far, the achievements in this field have been obtained in closed systems with fixed and definite rules, and the robustness is poor. Even the slightest unexpected change in the system environment may cause it to fail. Secondly, classical artificial intelligence is actually just a framework for building world cognitive models. This architecture cannot achieve cognitive collaboration. Specifically, this framework has not modeled the cognitive control process of vision and hearing. Jerry Fodor believes that artificial neural networks cannot support a complete human cognitive architecture. This way of simulating human rational cognitive representation results not only consumes huge human and material resources but is also difficult to fully complete tasks. Although scanning technology is getting better every year, it is still not enough to solve the mystery of abstract thinking in the brain. In other words, the investment in neuroscience research is large, but the research prospects are uncertain. In this case, if we put all the "eggs" of AI research in the "basket" of neuroscience research, there is a certain risk. However, at present, humans are in the era of weak artificial intelligence for general artificial intelligence, and superconducting quantum computing technology has theoretically given the possibility of achieving general artificial intelligence. The tremendous acceleration of computing power, especially the rise of quantum computers in recent years, has forced scientists to think about how to build cognitive models on the basis of quantum computers different from traditional Turing machine computing models.

Developing a unified theory and cognitive architecture for AGI requires breaking traditional knowledge-centered approaches, developing a universal methodology, understanding the principles behind deep learning, and distinguishing between different paradigms in computer science and neuroscience. This will ultimately lead to breakthroughs in scientific applications of AI. In summary, the pursuit of a unified theory and cognitive architecture for artificial intelligence involves the integration of six key areas: computer vision, natural language processing, machine learning, cognitive reasoning, robotics, and multi-agent systems. Traditional knowledge-centered approaches have been replaced by machine learning, which learns from data rather than relying on hand-coded knowledge. To advance the development of general artificial intelligence, we need to establish rules and learning mechanisms that break the current "black box" algorithms of deep learning and develop a universal methodology applicable to different domains and scenarios. This requires understanding the dependence of deep learning models on hyperparameters, the principles behind deep learning, and the development of robust and interpretable methods that are less reliant on large amounts of annotated data. Additionally, we need to develop databases and training platforms for next-generation AI methods and infrastructure. There are two main approaches to developing AGI: one is computer science-based, which simulates human behavior through machine programming without exploring the underlying driving force; the other is neuroscience-based, which aims to understand the internal driving force behind human behavior and design an artificial neural network to simulate these neuro-psychological functions. The mathematical methods and paradigms required for these approaches differ. Therefore, the mathematical foundation for AGI is a third approach or research method beyond computer science and neuroscience, which different AI algorithms can support each other through a common representation, ultimately leading to breakthroughs in scientific applications of AI. The core of this approach is distinguishing between the arithmetic paradigm of computers, the statistical paradigm of learning machines, and the geometric paradigm of understanding machines, all of which are tied together by a logical paradigm.

**3. Humanism in the Development of Artificial Intelligence**

Humans are the subject of scientific activity, both as discoverers and creators, as well as the ultimate beneficiaries of technological advancements. Artificial intelligence, as a human-made product of the information age, is a creation of humans themselves. Scientists should not only focus on instrumental rationality, but also possess value rationality. Human-centeredness is considered a core aspect of AI development and governance. Enhancing the social sustainability of AI deployment is essential. Humanism is a philosophical and ethical stance that emphasizes the value and agency of human individuals and collectives. From a human-centered perspective, values such as empathy, compassion, creativity, and social responsibility are crucial for the development of AGI. The integration of humanistic values in AGI development aims to create machines that can function as responsible and ethical members of society, rather than just tools for human use.

Integrating Ethics and Values in AGI, the human-centered perspective emphasizes the importance of human values, beliefs, and dignity in all aspects of life, including technological development. As the capabilities and potential of artificial intelligence systems continue to grow in the context of AGI development, this perspective becomes increasingly important. The human-centered view asserts that the development of AGI must reflect human values and priorities and consider the ethical and societal impacts of artificial intelligence. Developing AGI from a human-centered perspective involves designing AI systems that are not only technologically advanced but also considerate of the human experience. This includes developing AI systems that can understand and respond to human emotions, preferences, and values.

The human-centered view also emphasizes the need for transparency, accountability, and ethical considerations in the development of AGI systems. This requires a comprehensive approach that incorporates the perspectives of ethicists, policymakers, and other stakeholders. Overall, the human-centered view is crucial for the development of technically advanced and ethically responsible AGI. Therefore, this perspective must be considered throughout the entire process of AGI development, from research and design to deployment and regulation.

The importance of prioritizing human needs, values, and experiences in the development and governance of artificial intelligence, with human-centeredness as the core concept and objective. The motto of Stanford University's Human-Centered Artificial Intelligence (HAI) lab is "putting humans at the center," which means prioritizing human needs, values, and experiences in specific contexts. In the development and governance of artificial intelligence, the human-centered approach aims to ensure that technology is designed and used for the betterment of human well-being and society as a whole. Human-centeredness is considered the core of AI development and governance. The shortcomings of Human-Centered AI (HCAI) lie in not specifying how to achieve it. Human-centeredness explains the important role and necessity of AI in technological liberation, and HCAI has become a key concept and objective in policy documents aimed at developing public governance of AI. Its commitment to setting governance goals and practices is to enhance human performance and support better human-technology interaction, as well as the commitment of public governance institutions to guide technology development and deployment to serve and benefit humanity.

The impact of human-centered artificial intelligence on society and individuals. The article discusses the impact of human-centered artificial intelligence (HCAI) on society and individuals. While HCAI has the potential to improve people's lives, enhance public services, and increase human performance, there are also concerns about the risks involved and who benefits from it. It is important to conduct empirical research to understand the effects of HCAI and to develop frameworks, processes, and tools to promote cooperation and establish common values. The article suggests that the use of HCAI in policy documents may undermine its potential to promote human welfare and common interests, and calls for reflection and reform in its application.

In China, people-oriented development is a trait of human practice guided by Marxism, which emphasizes that the fundamental purpose and value orientation of social development is the interests and happiness of the people. People-oriented development is a philosophical and ethical stance that stresses human dignity, value, and agency, advocating for human well-being, freedom, and autonomy. From a people-oriented perspective, the development of AGI should serve the interests and needs of humans, respect their rights and values, and contribute to their prosperity and happiness. However, incorporating people-oriented values into AI development faces various challenges, such as the lack of a shared understanding of what those values are and how to implement them in AI design. Moreover, integrating people-oriented values into AI development has raised ethical and social issues, such as unintended biases and discrimination in decision-making, negative impacts on employment and privacy, and concerns about the fair distribution of benefits and harms among different social groups. Despite these challenges, China has been striving to develop AI systems that integrate people-oriented values, emphasizing the importance of developing "safe, controllable, and beneficial to society" AI in the "New Generation Artificial Intelligence Development Plan" released in 2017. This plan also stresses the necessity of interdisciplinary and cross-cultural collaboration to ensure that AI development aligns with people-oriented values. A people-oriented AI development framework proposed by Zhu and He (2019) considers ethical, cultural, and social factors. To achieve technological liberation, public governance and design of AI must be socially sustainable, placing the impacts on society, economy, and environment, as well as the needs and values of people and communities, at the center of AI governance and deployment.

HCAI has become an important concept and goal in research and policy documents on how to guide and design artificial intelligence to support the beneficial aspects of AI for individuals and society. In general, there are two approaches to human-centered AI: one originates from user-centered technology design, and the other represents its use in policy documents. From a user-centered design perspective (also known as human-centered design, or HCD), HCAI seeks to improve human-technology interaction and human performance by focusing on actual human capabilities, needs, and values, and ensuring that ethical principles are met in AI product design. Ethical principles in AI governance such as fairness, accountability, explainability, and transparency are central to this concept. The concept is more widely used in national AI strategies and policy documents, where human-centeredness has become a core but multi-voiced concept primarily used to bundle a set of moral and human rights principles as the foundation for AI strategies, objectives, or visions. One of the main documents in the EU's AI strategy, the EU's "Ethics Guidelines for Trustworthy AI" , states that AI systems "need to be human-centric, designed and developed with the aim of serving humans and the common good, with the goal of improving human well-being and freedom". In addition, it defines human-centered AI as a way of "ensuring that respect for fundamental rights is paramount in the development, deployment, use and monitoring of AI systems". In contrast to the user-centered design perspective (e.g., Shneiderman, 2022), the EU's ethical guidelines seem to broaden the goals of human-centered AI from improving human performance to serving the common good, enhancing well-being, and respecting human and fundamental rights. This perspective, which we call the emancipatory view, changes the ambition and expected impact that this concept carries. We identify three issues with the use of HCAI in policy documents from an emancipatory perspective.

**4. Governance of AGI from a human-centric perspective**

As the development of artificial general intelligence (AGI) continues to progress, there is a growing need to address the issue of governance from a human-centric perspective. Human-

Centered AI (HCAI) has emerged as an important concept and goal in research and policy documents on how to guide and design AI to support beneficial outcomes for individuals and society (Salo-Pontinen & Saariluoma, 2022; Shneiderman, 2022). There are two approaches to human-centric AI: one is rooted in user-centered technology design, and the other represents its use in policy documents. From the perspective of user-centered design, also known as Human-Centered Design (HCD), HCAI seeks to improve the interaction between humans and technology and human performance by focusing on actual human capabilities, needs, and values, and ensuring that ethical principles are adhered to in AI product design. Ethical principles such as fairness, accountability, explainability, and transparency in AI governance (Riedl, 2019; Leprince-Ringuet et al., 2021; Shneiderman, 2022). This concept is more broadly used in national AI strategies and policy documents, where human-centric has become a core but multi-voiced concept, primarily used to bundle a set of ethical and human rights principles as the foundation of AI strategies, goals, or visions (Salo-Pontinen & Saariluoma, 2022). governance of AGI from a human-centric perspective seeks to ensure that the development and use of AI systems are aligned with human values, needs, and principles. It is essential to consider the ethical implications of AI and prioritize the well-being of individuals and society when designing and deploying AI systems.

**4.1 Considerations for sustainable and inclusive governance of AI**

The United Nations (2012) emphasizes the need to consider the systemic impacts of technology on humans and society when focusing on sustainability. This requires an understanding of the complex and dynamic interactions between technology, operators, users, citizens, and society. It also means considering how these intelligent agents are influenced by technology and how the new AI culture is changing the human ecosystem. Furthermore, short-term and long-term, direct and indirect economic, social, and environmental opportunities, problems, and risks associated with the development and deployment of AI systems must be taken into account.

With regards to governance procedures, this perspective implies considering who and what viewpoints constitute development strategies and practices, as well as the extent to which this process is inclusive, democratic, and sustainable. To be effective, calls for sustainable governance and inclusive decision-making require a certain degree of mutual trust, communication, and transparency (Buhmann and Fieseler, 2022; Stahl, 2022). This is achieved by ensuring that the technical and organizational practices of data and service ecosystems are aligned with safety and ethical standards.

One of the main documents of the European Union's AI strategy, the EU's Ethics Guidelines for Trustworthy AI (AI HLEG, 2019, p.4) states that AI systems "need to be human-centric, committed to serving humans and the common good, with the goal of improving human well-being and freedom." In addition, it defines human-centric AI as a way of "crucially ensuring that human values guide the development, deployment, use, and monitoring of AI systems by respecting fundamental rights" (AI HLEG, 2019, p.37). However, the EU's ethical guidelines seem to broaden the goal of human-centric AI from improving human performance to serving the common good, improving well-being, and respecting human rights and fundamental rights, unlike the user-centered technology design perspective (e.g., Shneiderman, 2022). This perspective, which we refer to as the liberation view, changes the ambition and expected impact carried by this concept. We identify three issues with the use of HCAI in policy documents in a liberation view. The possibility of maintaining and building secure and robust AI systems, as well as the transparency and interpretability of AI models, can enable accountability, which is a necessary element for establishing trust in the deployment and development of AI by organizations and industries (AI HLEG, 2019; Sutrop, 2019; Schneiderman, 2022).

**4.2 Trust Frameworks for Governance in AGI Development**

Concept of building trust in technology governance was first introduced at the Global Future Council (GFC) by the World Economic Forum in 2016. Since then, trust has been considered a fundamental element in using artificial intelligence (AI) by various stakeholders. Trust in AI includes credible research, trustworthy AI designers and developers, reliable organizations, trustworthy design principles and algorithms, and responsible AI application deployment. Establishing and maintaining trust involves both social and technological structures and ensuring accountability of AI systems even in complex usage environments. The development of a trustworthy AI society requires principles of managing communication activities and a reliable governance system to support democratic processes and transparency policies when addressing social challenges associated with AI use. The article highlights that understanding the prerequisites of trust is essential for developing social rules for deploying AI, as trust is critical in building social capital that binds society together and is a prerequisite for sustainable data economy and AI use. The article concludes that a multi-level governance challenge exists, and public administrative departments need to develop new practical governance frameworks and tools to support the challenges, solutions, and values that guide AI use and development.

Towards Human-Centered Paradigm Shift in AI Development. The shift towards a human-centered paradigm is necessary to establish trust in AI deployment for society. A systemic approach that accepts and implements human-centered liberation goals is required in the current AI governance mechanisms. This approach can be supported by novel software solutions to enhance the design, deliberation, and collaboration processes. The systemic approach needs to rethink and creatively adjust processes and technologies to enable engagement and design when public AI governance needs to be sufficiently flexible to adapt to complex and dynamically changing circumstances. It requires fostering enlightened communication, reciprocity, inclusivity, and transparency in integrating the dynamic processes of society, citizens, and various data, services, and earth ecosystems, all of which are encompassed by the ethical, sustainable, and trustworthy AI principles outlined in this article.

HCAI has become an important concept and goal in research and political papers that focus on guiding and designing AI to support the beneficial aspects of AI for individuals and society. The EU's "Ethics Guidelines for Trustworthy AI" (AI HLEG, 2019, p. 4) states that AI systems "need to be human-centric, devoted to serving humans and the common good, with the goal of improving human welfare and freedom." Another approach to developing AI from a human-centered perspective is to combine human-like intelligence and emotions in AI systems. This approach involves developing AI systems that can simulate human-like decision-making processes, emotions, and social skills. Such systems can be designed to interact with humans in a more natural and empathetic way, enhancing the user experience and increasing societal acceptance of AI.

Trust Framework for AGI from Three Perspectives explores a trust framework for artificial general intelligence (AGI) from three perspectives: (1) design principles. Design principles should ensure that human values and moral beliefs are consistent with the values embodied in AGI design. AGI should be interpretable, which is the foundation for building trust between humans and machines, ensuring accountability and transparency of system use. AGI should be designed to collaborate with humans, emphasizing the integration of human and machine abilities and innovative education models, rather than replacing humans.

**(1) Human-Centered Design in AI Development**

Human-centered design (HCD) is a method that prioritizes the needs and experiences of human users in the design process. It involves empathizing with users to understand their needs, defining problems, ideating solutions, prototyping and testing, and iterating until the final product meets user needs. This design method is applicable in various fields, including the development of artificial intelligence (AI) systems.The development of AI systems requires human-centered design methods. This is because AI systems not only need to solve technical problems but also need to consider the needs and experiences of end-users. Human-centered design methods can ensure that the design of technology considers end-users, making it more likely to be accepted and effective.

There are many points of connection between human-centered design methods and the development of AI systems. For example, AI systems need interpretability to establish trust between humans and machines and ensure accountability and transparency in their use. These issues need to be considered during the design stage to ensure that the final product can effectively serve users.During the development of AI systems, designers need to work closely with end-users to ensure their needs are fully met. This requires designers to understand user behavior patterns, needs, and expectations to better consider these factors in the design process. Therefore, designers need to actively communicate with users throughout the development cycle and adapt to their feedback and suggestions.In the human-centered design process, prototyping and testing are essential. Prototyping can help designers and users better understand the system's functions and interface. Testing can help designers understand the system's efficiency and usability and identify potential problems. Therefore, designers should conduct multiple rounds of prototyping and testing to ensure that the final product meets user needs when developing AI systems.

In addition, human-centered design methods also need to consider the sustainability and scalability of AI systems. This means that designers need to consider the system's lifecycle and ensure that it can be expanded and updated in the future. This requires designers to use sustainable development methods and ensure that the system's code and architecture are scalable.In summary, human-centered design methods are essential in the development of AI systems. This method can ensure that the system's design considers end-users and has interpretability, accountability, and transparency. When developing AI systems, designers should work closely with end-users, conduct multiple rounds of prototyping and testing, and consider the system's sustainability and scalability.

**(2) Ethical and Responsible AI**

As AI systems become increasingly advanced and widespread, it is increasingly important to ensure they are developed and used in an ethical and responsible manner. This includes considerations such as fairness, accountability, transparency, privacy, and security. Ethical and responsible AI frameworks provide guiding principles for the development and deployment of AI systems that align with human-centered values. To address the ethical and societal impacts of AI development, it is important to consider responsible AI principles. Responsible AI refers to the development and deployment of AI systems that align with ethical and societal values while minimizing harm and maximizing benefits to all stakeholders. Some key principles of responsible AI include transparency, fairness, accountability, privacy, and human oversight.In conclusion, as AI technology continues to advance, it is essential to ensure that it is developed and used in an ethical and responsible manner. Ethical and responsible AI frameworks provide guiding principles for the development and deployment of AI systems that align with human-centered values. By considering factors such as fairness, accountability, transparency, privacy, and human oversight, we can ensure that AI systems are developed and used in ways that maximize benefits to all stakeholders while minimizing harm.()

**(3) Human-AI Collaboration**

Artificial General Intelligence (AGI) refers to an AI system that has the ability to match or surpass human intelligence and has broader applications compared to current AI technologies. The development of AGI has the potential to have significant impacts on multiple aspects of society, economy, and ethics.A key issue related to AGI is human-AI collaboration, which refers to the process of humans and AI systems working together to accomplish a task. In human-AI collaboration, humans and AI systems have different roles and can complement and support each other to achieve common goals.Human-AI collaboration can take various forms. One common form is the collaboration between humans and AI systems to solve complex problems. For example, in the healthcare field, AI systems can analyze large amounts of medical data and provide diagnostic suggestions, while doctors combine their experience and judgment to make final diagnostic decisions. In this case, humans and AI systems can leverage their respective strengths to work together and improve diagnostic accuracy and treatment effectiveness.

Another form of human-AI collaboration is based on complementarity. In this case, AI systems can perform repetitive or low-level tasks, allowing humans to focus on higher-level tasks. For example, in the manufacturing industry, robots can perform repetitive assembly tasks while workers can focus on tasks that require higher-level skills.Human-AI collaboration can also be achieved through human supervision and control of AI systems. In this case, AI systems can provide assistance and support to humans but still require humans to make final decisions. For example, in autonomous driving, AI systems can provide navigation and control, but the ultimate driving decision is still made by the human driver.Although human-AI collaboration can bring many benefits, it also poses risks and challenges. For example, if the design of AI systems is inappropriate or not adequately supervised and controlled, it may lead to issues such as bias, discrimination, errors, or abuse. Additionally, human-AI collaboration may lead to inconsistent workflows and insufficient training, which can reduce the efficiency and quality of the system.Therefore, in human-AI collaboration, it is important to ensure that the design and deployment of AI systems align with ethical and responsible principles.

**(4) Explainable Artificial Intelligence (XAI)**

Explainable AI (XAI) refers to the ability of AI systems to provide transparent and interpretable explanations for their decisions and actions. XAI is important because it enables humans to understand how and why AI systems make specific decisions, which is crucial for ensuring that the technology aligns with human values and expectations. The concept of explainable artificial intelligence (XAI) has gained increasing attention in recent years. XAI refers to the ability of artificial intelligence systems to explain their decisions and behaviors in a way that is understandable to humans. This is crucial for ensuring transparency and accountability in AI systems, as well as building trust between humans and AI.

**(5) Value alignment of AGI in scientific activities**

Value alignment in scientific activities refers to the process of ensuring that the goals and purposes of AGI systems are consistent with human values and objectives. This involves identifying and prioritizing the most important human values and designing AGI systems that are consistent with those values. Value alignment is crucial to ensuring that the development and use of AGI systems are beneficial to humans and do not cause harm or unintended consequences.AGI, or artificial general intelligence, has the potential to profoundly impact scientific activities as it may help us better understand the nature of the universe and the natural world. However, as with any new technology, we must ensure that it aligns with human values and goals. This is known as value alignment, a process that ensures we identify and consider the most important human values to design AGI systems that are consistent with these values.

The importance of value alignment lies in the fact that if AGI systems are not aligned with human values, they may cause unintended consequences or harm, and even go against our will. Therefore, we need to ensure that our AGI systems are human-centered, to ensure they are beneficial to humans at all times and do not cause harm. The key to achieving this is to prioritize the most important human values and ensure that our AGI systems are designed to be consistent with these values.Through value alignment, we can ensure that AGI systems have a positive impact on scientific activities. For example, we can use AGI systems to better understand and explain phenomena in the natural world, thereby advancing scientific research. Additionally, AGI systems can help us optimize experimental design and data analysis, improving the efficiency and accuracy of scientific research. Throughout this process, we must ensure that our AGI systems are based on the right values and goals to ensure that they truly serve the interests of humanity.

**Conclusion**

The text discusses the importance of a human-centered approach to the development and governance of artificial intelligence (AI). This approach emphasizes the need to prioritize human values, beliefs, and dignity in all aspects of AI development and to consider the ethical and societal impacts of AI. It also stresses the importance of transparency, accountability, and ethical considerations in the development of AI systems. The text discusses the impact of human-centered AI on society and individuals, including both potential benefits and risks. It also highlights the challenges of incorporating human values into AI development and governance and proposes strategies for achieving this goal. The text concludes by emphasizing the importance of a socially sustainable approach to AI governance and deployment that places the needs and values of people and communities at the center of AI development.In general, AGI must be designed and used in a responsible and ethical way that respects human dignity, privacy, and autonomy. Developing AGI with a human-centered approach is of great significance for the future of technology and society in China. Integrating human-centered values such as empathy, ethics, and social responsibility into the development and deployment of AGI can lead to more ethical and sustainable use of the technology.

At the policy level, strengthening international cooperation for AGI and promoting the sharing of research results is fundamental to improving the ability to respond to emergencies and ensuring the successful implementation and expansion of AGI. Advanced AI research centers are located in the United States, Canada (especially Toronto and Montreal), Europe (London, Paris, and Berlin), and Israel. For example, 25 European countries signed the "AI Cooperation Declaration," committing to cooperation, dialogue, and seeking consensus on AI research and application cooperation between countries. Strengthening international cooperation for AGI and promoting the sharing of research results is fundamental to improving the ability to respond to emergencies and ensuring the successful implementation and expansion of AGI. The importance of international cooperation for AGI is already highly valued, and some countries and regions are providing policy support for international cooperation through legislation. For example, the US government established the AI Security Commission, and the European Union set up the High-Level Expert

Group on Artificial Intelligence to strive for the right to speak and rule-making power in technology development.

In terms of talent development, the scale, speed, and quality of talent cultivation for AGI cannot meet the needs of field development, especially for local talent. It is urgent to strengthen talent cultivation for strong AI, which is the foundation of scientific research. With global leading AI concepts and grand goals as a guide, in the technical field, optimizing the mechanisms and environments for talent education, cultivation, and growth can rapidly develop a group of professionals with specialized research and development knowledge. In the management field, emphasis should be placed on cultivating entrepreneurs and operators who reflect commercial promotion and demand expansion characteristics. Through collaborative innovation among industry, academia, research, and application, a solid talent guarantee can be provided for the healthy and stable development of AGI. The interdisciplinary approach of liberal arts, medical science, engineering, and science can be promoted to cultivate "general, intelligent, and versatile" talents. In this approach, the neural operation mechanism of the biological brain and cognitive behavior mechanism become essential research areas, which can promote the further development of AGI technology.